\documentclass{svjour3}                     
\smartqed  
\usepackage{color,graphicx,hyperref,amsmath,amssymb,mathrsfs}

\begin{document}

\title{Reconstructing bifurcation behavior of a nonlinear dynamical
system by introducing weak noise}
\titlerunning{Reconstructing bifurcation behavior of a nonlinear dynamical
system}
\author{Debraj Das \and Sayan Roy \and Shamik Gupta}
\institute{Debraj Das (Equal Contribution) \at Department of Physics, Ramakrishna Mission Vivekananda University, Belur Math, Howrah 711202, India
\and
Sayan Roy (Equal Contribution) \at Department of Physics, Indian Institute of Science
Education and Research Bhopal, Bhopal Bypass Road, Bhauri, Bhopal 462
066, Madhya Pradesh, India
\and
Shamik Gupta \at Department of Physics, Ramakrishna Mission Vivekananda
University, Belur Math, Howrah 711202, India.
\email{shamik.gupta@rkmvu.ac.in}}
\maketitle
\begin{abstract}
For a model nonlinear dynamical system, we show how one may obtain its
bifurcation behavior by introducing noise into the dynamics and then
studying the resulting Langevin dynamics in the weak-noise limit. A
        suitable quantity to capture the
bifurcation behavior in the noisy dynamics is the conditional
probability to observe a microscopic configuration at one time,
conditioned on the observation of a given configuration at an earlier
time. For our model system, this conditional probability is studied by
using two complementary approaches, the Fokker-Planck and the
path-integral approach. The latter has the advantage of yielding
exact closed-form expressions for the conditional probability. All our
predictions are in excellent agreement with direct numerical integration of the
dynamical equations of motion.
\end{abstract}

\section{Introduction}
\label{sec:intro}

Nonlinear dynamical systems present a plethora of physical phenomena
that are truly fascinating, but which at the same time appear
counterintuitive and intriguing, especially when viewed from
the perspective of linear
systems that are much simpler to understand and analyze~\cite{Strogatz:2014,Lakshmanan:2003}.
As examples, one may cite chaos~\cite{Ott:2002}, pattern
formation~\cite{Cross:2009}, solitons~\cite{Dauxois:2010}, and many
more. Despite the
intricacies and roadblocks involved in providing an analytical
characterization, nonlinear phenomena have attracted the attention
of physicists, engineers, biologists and mathematicians, a reason being
that nature is inherently nonlinear. 

A very interesting dynamical feature exhibited by nonlinear systems is
that of bifurcation, whereby a given dynamics exhibits qualitatively
different flow structure as one or more dynamical parameters are varied.
A consequence is that fixed points into which the dynamical variables settle at long times
may have different stability properties for different parameter ranges,
or they may even by created or destroyed as the dynamical parameters are
tuned across critical values. 

A deterministic dynamical system is typically characterized in terms of behavior of
specific initial conditions under the dynamical evolution. In contrast,
introducing noise into the dynamics requires a statistical description
in the form of a suitable distribution of the dynamical variables and a
study of its
evolution in time. In this work, we address the issue of how one may obtain the bifurcation
diagram of a nonlinear dynamical system by introducing noise into its
dynamics and studying the resulting noisy dynamics using tools of stochastic processes. We
show that a suitable quantity to capture the
bifurcation behavior in the noisy dynamics is the conditional
probability to observe a microscopic configuration of the dynamical
variables at one time, conditioned on the observation of a given configuration at an earlier
time. We study this conditional probability by two complementary approaches, the Fokker-Planck and the
path-integral approach, with the latter offering the advantage of yielding an exact closed-form
expression for the conditional probability. Our results demonstrate that
when considered in the limit of weak noise, the noisy dynamics is able
to reproduce the bifurcation diagram of the
noiseless dynamics. Such a conclusion may not seem very surprising in
retrospect, especially since in the weak-noise limit, the noisy dynamical
trajectories represent small fluctuations about those for the noiseless
one. Our work primarily serves as a proposal of a theoretical framework to systematically
obtain the stability properties of the noiseless dynamics from a
suitable analysis of the noisy one, and as an illustration of how one may
derive analytical
expressions of the quantities involved in the latter analysis.  

The paper is laid out as follows. In Section~\ref{sec:model}, we
present our model system described in terms of noiseless time evolution of a
single phase-like variable on a potential landscape. We discuss some of
the dynamical
features of the system, and also introduce its noisy variant involving
time evolution in presence of Gaussian, white noise. An analysis of the bifurcation behavior
of the noiseless dynamics is taken up in Section~\ref{sec:analysis-noiseless-dynamics}. The noisy
dynamics is studied in Section~\ref{sec:analysis-noisy-dynamics} using two independent approaches, the Fokker-Planck and the path-integral
approach. In Section~\ref{sec:results}, the results obtained in the noisy dynamics in the limit of
weak-noise are compared with those for the noiseless dynamics,
allowing us to demonstrate how our objective of obtaining the
bifurcation diagram of the noiseless dynamics from the noisy one is
achieved. The paper ends with conclusions in
Section~\ref{sec:conclusions}.  

\section{The model}
\label{sec:model}
We consider a dynamical system described by a single phase-like variable
$\theta \in [-\pi,\pi]$, whose time evolution is given by 
\begin{equation}
\frac{{\rm d}\theta}{{\rm d}t}=A \sin \theta-B\sin 2\theta.
\label{eq:eom-D0-0}
\end{equation}
Here, the dynamical parameters $A$ and $B$ are real constants.
One may get rid of one of the parameters from the dynamics by a simple rescaling of time,
so that from now on we will consider the dynamics
\begin{equation}
\frac{{\rm d}\theta}{{\rm d}t}=a\sin \theta-\sin 2\theta,
\label{eq:eom-D0}
\end{equation}
where $a$ is a real constant.

The noisy dynamics corresponding to the noiseless evolution~(\ref{eq:eom-D0}) is obtained by introducing a Gaussian,
white noise term $\eta(t)$ on the right hand side of
Eq.~(\ref{eq:eom-D0}). One has consequently the
following Langevin dynamics:
\begin{equation}
\frac{{\rm d}\theta}{{\rm d}t}=a \sin \theta-\sin 2\theta+\eta(t),
\label{eq:eom}
\end{equation}
where the noise $\eta(t)$ satisfies 
\begin{equation}
\langle \eta(t)\rangle=0,~~\langle
\eta(t)\eta(t')\rangle=2D\delta(t-t'),
\label{eq:noise}
\end{equation}
with $D$ a positive constant, and angular brackets denoting average over noise realizations. Note that the parameter $D$ sets the strength of the noise,
and setting it to zero reduces the noisy dynamics to the noiseless one,
Eq.~(\ref{eq:eom-D0}).

Equation~(\ref{eq:eom}) corresponds to overdamped dynamics of $\theta$
in a potential $V(\theta)$, as
\begin{equation}
\frac{{\rm d}\theta}{{\rm d}t}=-V'(\theta)+\eta(t),
\label{eq:eom-with-noise}
\end{equation}
with
\begin{equation}
V(\theta)\equiv a \cos \theta-\frac{1}{2}\cos 2\theta,
\label{eq:Vtheta}
\end{equation}
and the prime denoting first derivative with respect to $\theta$.

\begin{figure}[!h]
\centering
\includegraphics[width=9cm]{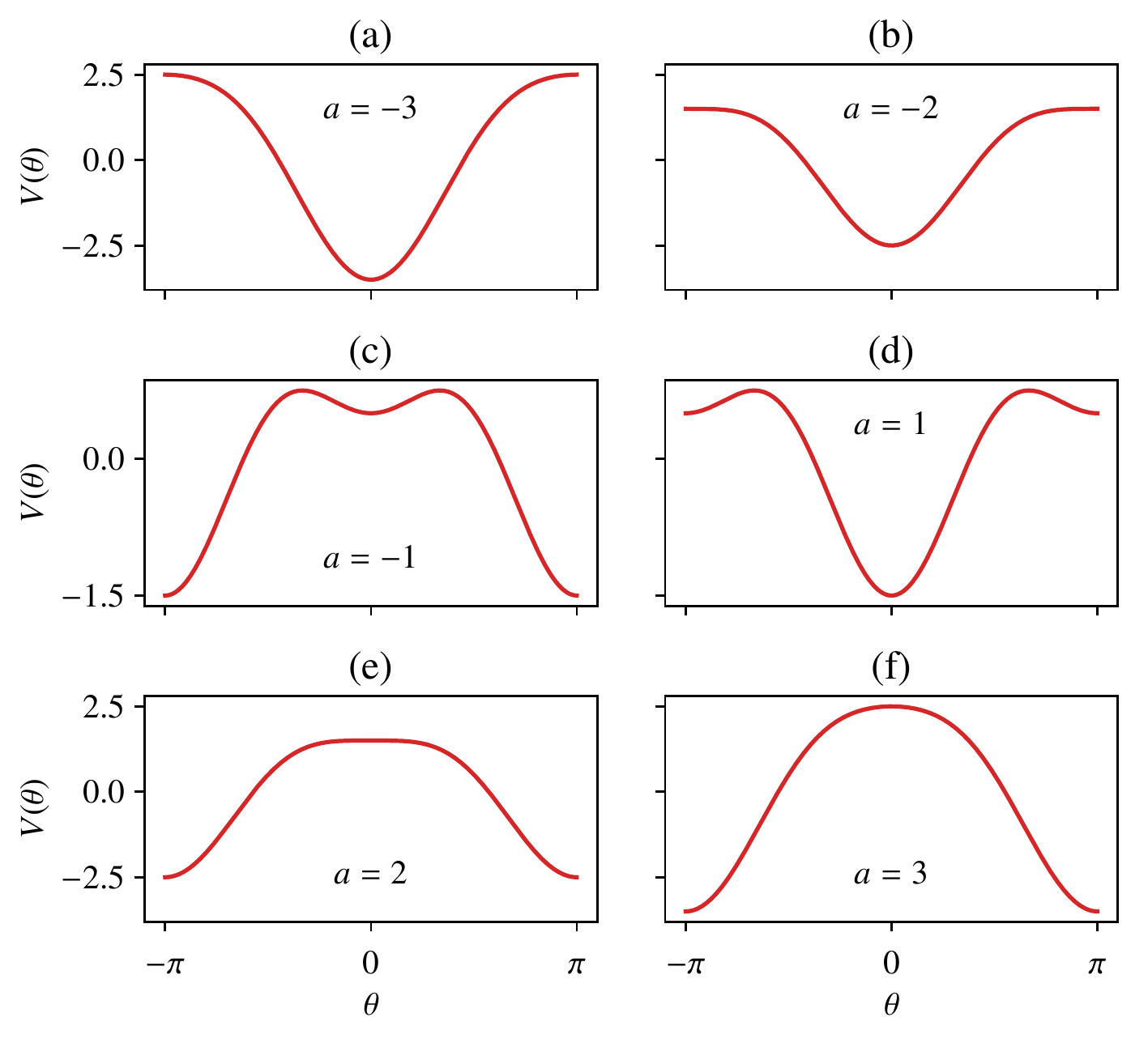}
\caption{(Color online) Potential $V(\theta)=a \cos \theta-(1/2)\cos 2\theta$ for representative values of the parameter $a$.}
\label{fig:Vtheta}
\end{figure}

Note that the potential satisfies $V_{a<0}(\theta)=V_{a>0}(\theta-\pi)$.
Solving $V'(\theta)=0$ gives for all $a$ the solutions $\theta=0,\pm
\pi$ as well as $\theta=\cos^{-1}(a/2),~\sin \theta=\pm \sqrt{1-a^2/4}$ for $-2 < a < 2$. It is
easily checked that $\theta=0$ is a maximum of $V(\theta)$ for $a>2$ and is a
minimum for $a<2$, while $\theta=\pm \pi$ maximize $V(\theta)$ for
$a<-2$ and minimize it for $a>-2$. Finally, $\theta=\cos^{-1}(a/2)$ is
a maximum for $-2 < a <2$. On the other hand, we have $V''(0)=0$ for $a=2$ and $V''(\pm \pi)=0$ for $a=-2$, while
$V''(\theta=\cos^{-1}(a/2);~-2< a < 2)=0$ for $a=\pm 2$.
Figure~\ref{fig:Vtheta} shows the potential $V(\theta)$ for
representative values of $a$.

\section{Analysis of the noiseless dynamics}
\label{sec:analysis-noiseless-dynamics}
The fixed
points $\theta^\star$ of the noiseless dynamics~(\ref{eq:eom-D0})
satisfy
$V'(\theta^\star)=0$, and hence are given by $\theta^\star=0,\pm \pi$ for
all values of $a$, with additional fixed points
$\theta^\star=\cos^{-1}(a/2),~\sin \theta^\star=\pm\sqrt{1-a^2/4}$ for
$a$ lying in the range $-2 < a < 2$. The linear stability of these fixed points may be determined
by substituting in Eq.~(\ref{eq:eom-D0}) the expansion
$\theta=\theta^\star+\Delta \theta$, with $|\Delta \theta|$
small, and keeping terms to linear order in $\Delta \theta$. One obtains
\begin{equation}
\frac{{\rm d}\Delta\theta}{{\rm d}t}=-V''(\theta^\star)\Delta \theta.
\end{equation}
It then follows that $V''(\theta^\star)>0$ (respectively,
$V''(\theta^\star)<0$) makes the perturbation
$\Delta \theta$ decay (respectively, grow) exponentially in time, rendering
$\theta^\star$ linearly stable (respectively, unstable). Using the properties of
$V(\theta)$ discussed earlier, we conclude that
\begin{itemize}
\item For $a>2$, the linearly stable fixed points are $\theta^\star=\pm \pi$,
while $\theta^\star=0$ is linearly unstable. For $a<-2$, the
stability of these fixed points gets exchanged.
\item For $a=2$, the fixed point $\theta^\star=0$ is linearly neutrally stable,
while $\theta^\star=\pm \pi$ are linearly stable.
For $a=-2$, the fixed points $\theta^\star=\pm \pi$ are linearly neutrally
stable, while $\theta^\star=0$ is linearly stable.
\item For $-2 < a <2$, the fixed points $\theta^\star=0,\pm \pi$ are
linearly stable, while $\theta^\star=\cos^{-1}(a/2);~\sin \theta^\star=\pm \sqrt{1-a^2/4}$ are linearly unstable. 
\end{itemize}

\begin{figure}[!ht]
\centering
\includegraphics[width=9cm]{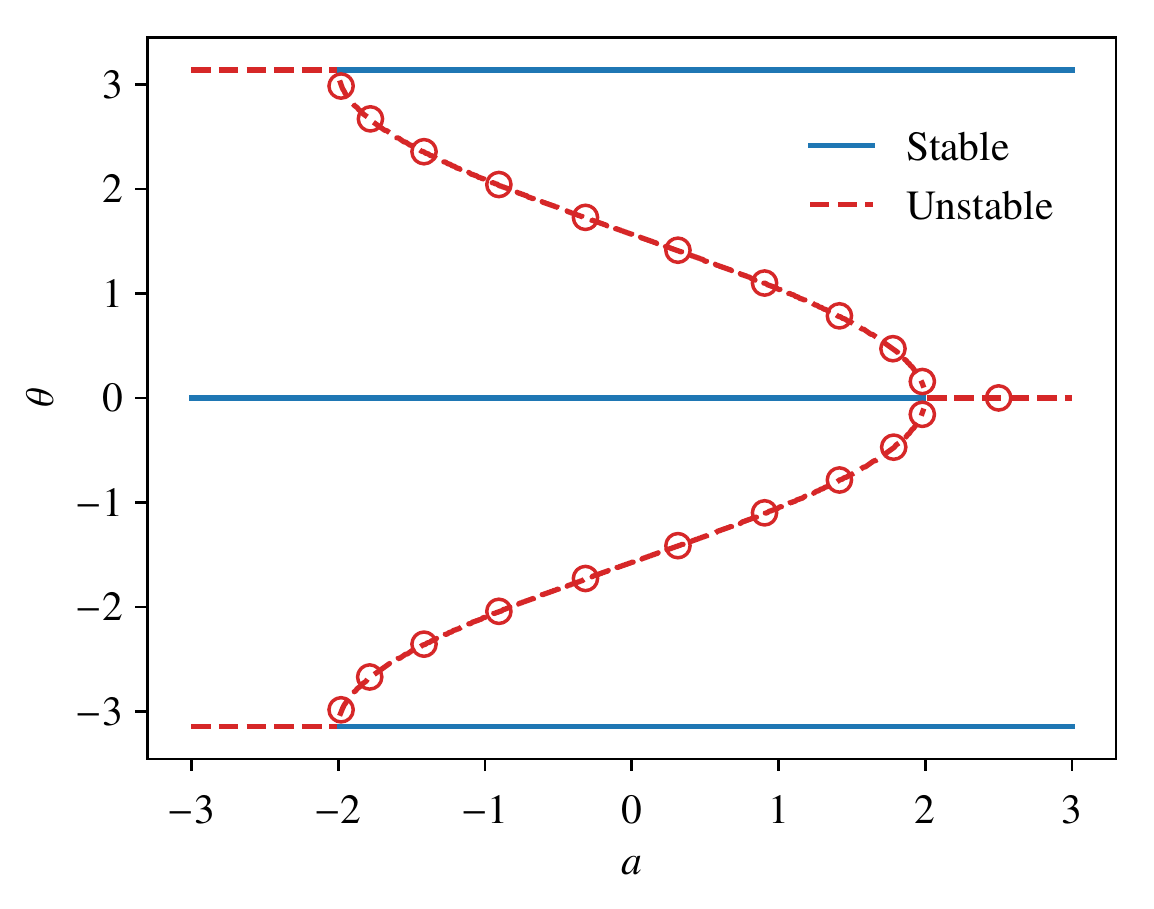}
        \caption{(Color online) Bifurcation diagram of the noiseless
        dynamics~(\ref{eq:eom-D0}). For all values of $a$, the fixed points are $\theta=0,\pm
        \pi$, while for $a$ in the range $-2 < a <2$, additional fixed
        points are given by $\cos \theta=(a/2)$; stable fixed points are
        denoted by continuous lines, while the unstable ones are denoted
        by dashed lines. Here, the
        red circles may be obtained from the exact analysis of the noisy
        dynamics (\ref{eq:eom}) based on the Fokker-Planck and the
        path-integral approach discussed in the text.}
	\label{fig:bifurcation}
\end{figure}

On the basis of the foregoing, one obtains the bifurcation diagram of
Fig.~\ref{fig:bifurcation} that shows the stable (continuous blue lines)
and unstable (dashed red lines) fixed
points as a function of $a$. Coexistence of multiple stable fixed points
for $-2 < a <2$ implies hysteretic behavior for the
model~(\ref{eq:eom-D0}). Let us identify a $2\pi$-periodic
variable of $\theta$ as a suitable order
parameter that captures this behavior. Since the stable fixed points are
either zero or $\pm \pi$, one may choose $\cos \theta$ as the simplest
such order parameter. 

\begin{figure}[!ht]
\centering
\includegraphics[width=9cm]{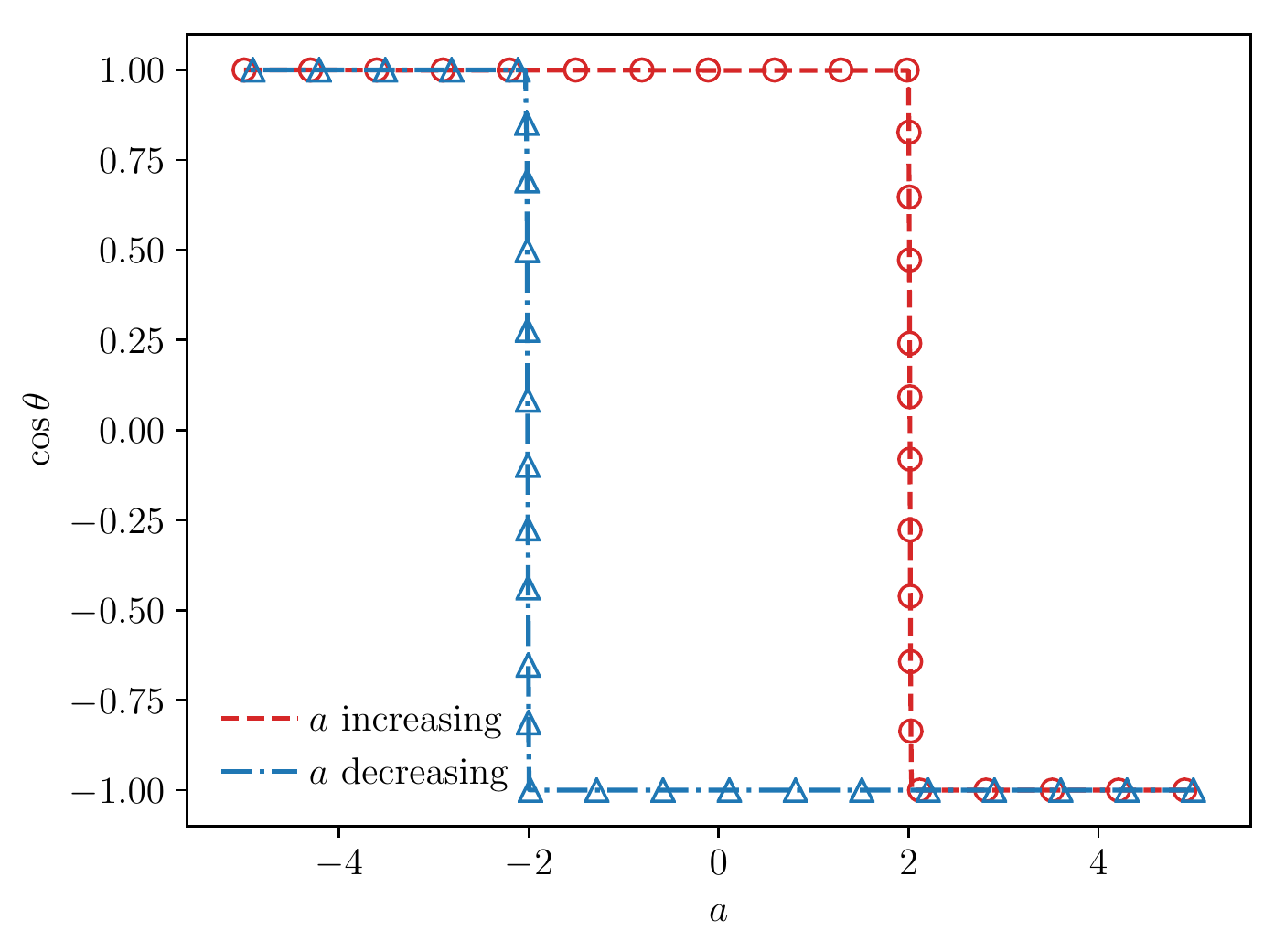}
\caption{(Color online) $\cos \theta$ as a function of
        adiabatically-tuned $a$, showing hysteretic behavior. The curves follow from the linear
        stability analysis of the noiseless dynamics~(\ref{eq:eom-D0})
        discussed in the text. On the other hand, the red circles and
        the blue triangles are obtained by numerically
        integrating the noisy dynamics~(\ref{eq:eom}) for the initial
        condition $\theta_0=0.75\pi$, and with $D=10^{-5}$ and time step ${\rm
        d}t=10^{-3}$; we first let the
        system reach the stationary state (reached at time $t=10$) at a
        given value of $a < -2$, and then
        increase $a$ adiabatically to high values and back in a cycle; Here, the data
        correspond to one realization of the noisy dynamics.} 
	\label{fig:hysteresis}
\end{figure}

We now obtain
the behavior of the stable value of $\cos \theta$ as $a$ is tuned
adiabatically from
small to large values and back. Adiabatic tuning of $a$ ensures that the
system while starting from an initial state has enough time to relax to
the stable state before the value of
$a$ changes appreciably. Referring to Fig.~\ref{fig:Vtheta}, if
one starts with a value of $a$ smaller than $-2$, any initial $\theta$
will relax at long times to the stable fixed point at $\theta^\star=0$.
As $a$ is now adiabatically tuned to higher values, the value of $\theta$ will remain
pinned to zero, until the minimum at $\theta=0$ of the
potential $V(\theta)$ turns into a maximum. The value of $a$ at which
this happens, obtained by solving $V''(0)=0$, is given by $a=2$.
Beyond $a=2$, the stable value of $\theta$ will change to the value at
the new minima, given by $\theta = \pm \pi$. Concomitant with the
aforementioned behavior, $\cos \theta$ versus $a$ will behave as shown
in Fig.~\ref{fig:hysteresis} for the case of increasing $a$. Following
the above line of argument, one may easily obtain the behavior of $\cos
\theta$ versus $a$ for the case when $a$
has a starting value greater than $2$ and is
adiabatically decreased to a value less than $-2$. The corresponding
behavior is depicted in Fig.~\ref{fig:hysteresis} for the case of
decreasing $a$. Hysteretic behavior of $\cos
\theta$ is clearly evident from the figure.

With respect to the bifurcation diagram~(\ref{fig:bifurcation}), one may
wonder about the nature of bifurcation at the point $(a=2,\theta=0)$: on
decreasing $a$ across $a=2$, a line of unstable fixed points bifurcates into two
lines of unstable fixed points that are symmetrically disposed about a line of stable
fixed points. Close to the bifurcation point, expanding Eq.~(\ref{eq:eom-D0}) to the first
two leading orders in $\theta$, one gets
\begin{eqnarray}
\frac{{\rm d}\theta}{{\rm
        d}t}=(a-2)\theta+(8-a)\frac{\theta^{3}}{6},
\end{eqnarray}
which has the form of the so-called subcritical pitchfork
bifurcation~\cite{Strogatz:2014}. Proceeding similarly, it is easy to
see that the bifurcation that occurs as $a$ is increased through
$(a=-2,\theta=\pm \pi)$ is also a subcritical pitchfork bifurcation.

\begin{figure}[!ht]
\centering
\includegraphics[width=9cm]{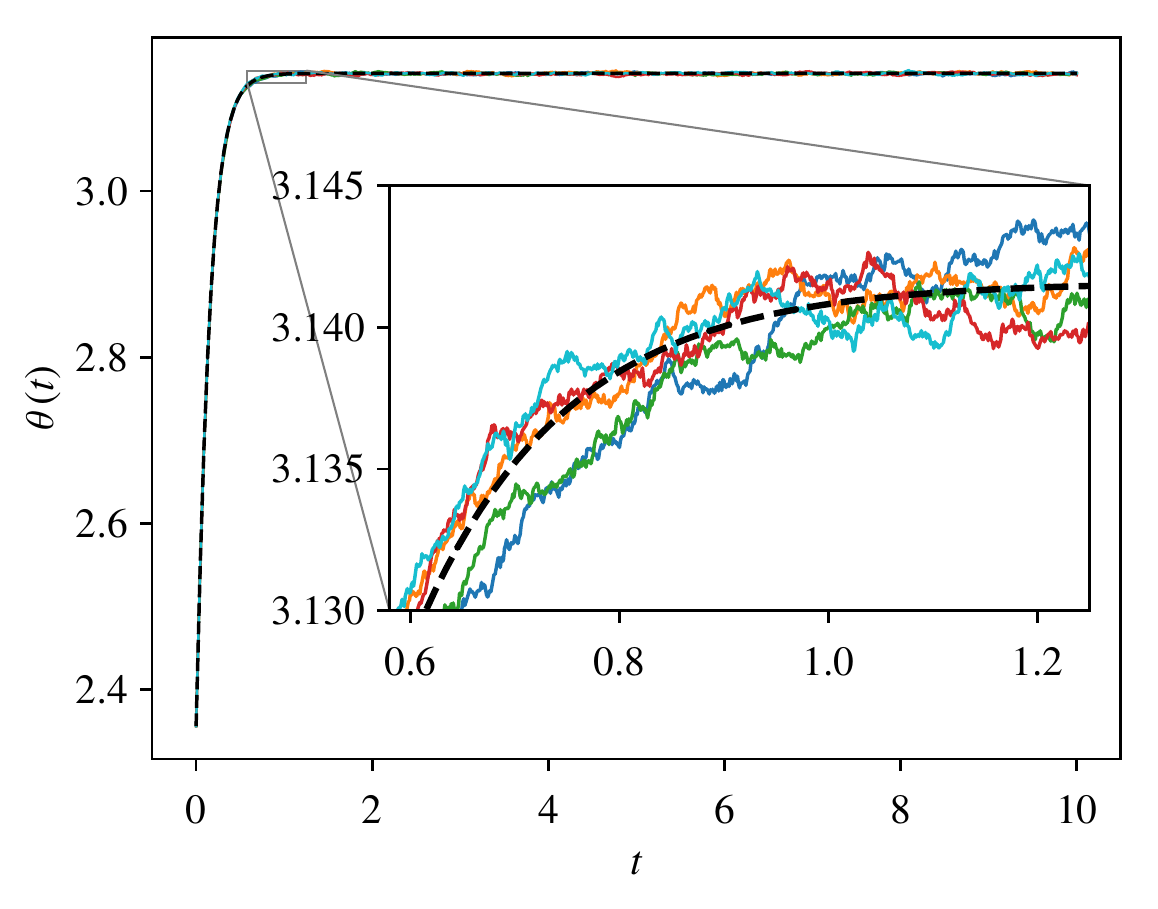}
\caption{(Color online) Dynamical trajectories of the noiseless and the
noisy dynamics, Eqs.~(\ref{eq:eom-D0}) and (\ref{eq:eom}), respectively, for the initial value
$\theta_0 = 0.75\pi$. The dashed line corresponds to the noiseless
case, while the five continuous lines correspond to five independent
realizations of the noisy dynamics. The trajectories are obtained by numerically integrating the
corresponding equation
of motion with integration time step ${\rm
d}t=10^{-3}$~\cite{note-euler}. Here, we have
chosen $a=5,D=10^{-5}$. One may observe that the trajectories for the noisy
case represent small fluctuations ($\sim \sqrt{D}$) about the 
noiseless trajectory.}
\label{fig:space-time-trajectory}
\end{figure}

Figure~\ref{fig:space-time-trajectory} shows the dynamical
trajectories for the noiseless and the noisy dynamics,
Eqs.~(\ref{eq:eom-D0}) and~(\ref{eq:eom}), respectively, from which one
may observe that in the weak-noise limit ($D \to 0$), the trajectories
for the noisy dynamics occur as small fluctuations ($O(\sqrt{D})$) about those for the
noiseless dynamics. This observation makes us anticipate that it should be possible
to extract the bifurcation behavior of the noiseless
dynamics~(\ref{eq:eom-D0}) from a suitable analysis of
the noisy dynamics~(\ref{eq:eom}). A straightforward numerical check of this
expectation is offered by performing numerical integration of the noisy
dynamics~(\ref{eq:eom}) for small noise strength, obtaining the
values
of $\langle \cos \theta \rangle$ as a function of adiabatically-tuned
$a$, and comparing with the results of the noiseless dynamics.
Figure~\ref{fig:hysteresis} indeed shows a match between the two results.
Our aim in this work is to explain this match on the basis of a
theoretical analysis of the noisy dynamics. We therefore turn to such an analysis
in the next section. 

\section{Analysis of the noisy dynamics}
\label{sec:analysis-noisy-dynamics}

If one has to locate dynamically the stable fixed points of the noiseless
dynamics~(\ref{eq:eom-D0}) for a given value of $a$, one needs to initiate the dynamics by
specifying an initial condition for $\theta$ and then let the dynamics run for a long time in
order that it relaxes to a stationary state. The latter would correspond to
the stable fixed points of the dynamics. In the case of noisy
dynamics~(\ref{eq:eom}), the system
while evolving from the same initial condition will have at long times a range of
possible values of $\theta$ corresponding to different dynamical
trajectories attained with different realizations of the noise
$\eta(t)$. In this case, it is then pertinent for an analytic
characterization of the dynamics that one defines a conditional
probability density $P(\theta,t|\theta_0,0)$, which gives the probability density that the phase has the value $\theta$
at time $t$, given that it had the value $\theta_0$ at the initial
instant $t=0$. Our expectation is that studying $P(\theta,t|\theta_0,0)$
as $t \to \infty$ and $D \to 0$ should
allow to recover the stable fixed points of the noiseless dynamics. 

The function $P(\theta,t|\theta_0,0)$ is $2\pi$-periodic in both
$\theta$ and $\theta_0$:
\begin{equation}
P(\theta+2\pi,t|\theta_0+2\pi,0)=P(\theta,t|\theta_0,0),
\end{equation}
and obeys the normalization 
\begin{equation}
\int_{-\pi}^\pi {\rm
d}\theta~P(\theta,t|\theta_0,0)=1~\forall~\theta_0,t.
\end{equation}

\subsection{The Fokker-Planck approach}
\label{sec:Fokker-Planck-equation}
In this subsection, we discuss how one may obtain for a given value of
$\theta_0$ the conditional probability density $P(\theta,t|\theta_0,0)$ as
a function of $t$ by solving the time evolution equation it satisfies. 
The time evolution of $P$ is given by a
Fokker-Planck equation that may be written down straightforwardly by using
the Langevin equation~(\ref{eq:eom-with-noise}). One gets
\begin{eqnarray}
\frac{\partial P(\theta,t|\theta_0,0)}{\partial
t}&=&-\frac{\partial }{\partial \theta} \left[ -V'(\theta)P(\theta,t|\theta_0,0) \right]+D\frac{\partial^2
P(\theta,t|\theta_0,0)}{\partial \theta^2},
\label{eq:Fokker-Planck}
\end{eqnarray}
with the initial condition
\begin{equation}
P(\theta,0|\theta_0,0)=\delta(\theta-\theta_0).
\label{eq:P-initial-condition}
\end{equation}

In order to solve Eq.~(\ref{eq:Fokker-Planck}), noting that $P$ is
$2\pi$ periodic in $\theta$, one may expand it in a
Fourier series in $\theta$: 
\begin{equation}
P(\theta,t|\theta_0,0)=\sum_{n=-\infty}^\infty
\widetilde{P}_{n}(t|\theta_0,0)e^{in\theta},
\end{equation}
with $P(\theta,t|\theta_0,0)$ being real implying that
$[\widetilde{P}_n(t|\theta_0,0)]^\star=\widetilde{P}_{-n}(t|\theta_0,0)$,
and star denoting complex conjugation.
Substituting in Eq.~(\ref{eq:Fokker-Planck}), one obtains the time
evolution of the Fourier coefficients $\widetilde{P}_n$ as
\begin{eqnarray}
&&\frac{\partial \widetilde{P}_{n}(t|\theta_0,0)}{\partial
t} = -Dn^2\widetilde{P}_{n}(t|\theta_0,0)\nonumber \\
&&+\frac{na}{2}\left[\widetilde{P}_{n+1}(t|\theta_0,0)-\widetilde{P}_{n-1}(t|\theta_0,0)\right]+
\frac{n}{2}\left[\widetilde{P}_{n-2}(t|\theta_0,0)-\widetilde{P}_{n+2}(t|\theta_0,0)\right],
\label{eq:eom-Fourier-coefficients}
\end{eqnarray}
with Eq.~(\ref{eq:P-initial-condition}) yielding 
\begin{equation}
\widetilde{P}_n(0|\theta_0,0)=\frac{1}{2\pi}e^{-in\theta_0}.
\label{eq:Ptilde-initial-condition}
\end{equation}

For any $n$, the system of coupled equations (\ref{eq:eom-Fourier-coefficients}) is not
closed and in fact involves an infinite hierarchy: for a given
value of $\theta_0$, to obtain
$\widetilde{P}_n(t|\theta_0,0)$ as a function of $t$ requires knowing $\widetilde{P}_{n+1}$ and
$\widetilde{P}_{n+2}$ whose
solution requires knowing $\widetilde{P}_{n+3}$ and $\widetilde{P}_{n+4}$, and so on.
For the initial condition~(\ref{eq:Ptilde-initial-condition}), however, the system of
equations may be solved easily by truncating it at a given value $n=n_{\rm
max}$, i.e., by stipulating that $\widetilde{P}_n(t|\theta_0,0)=0$
for $n > n_{\rm max}$ and for all $t$. Here, $n_{\rm max}$ may be chosen to be as large as possible.

\subsection{The path-integral approach}
\label{sec:path-integral-approach}
We now discuss a complementary approach to obtain $P(\theta,t|\theta_0,0)$ as a function of $t$, by invoking the
Feynman-Kac path-integral formalism of treating stochastic processes~
\cite{Feynman:2010,Schulman:1981,Kac:1949,Kac:1951}. An advantage is
that in contrast to the Fokker-Planck approach, one obtains in this
approach a closed-form expression for $P(\theta,t|\theta_0,0)$. To this
end, we \textbf{follow the general procedure discussed in
Ref.~\cite{Roldan:2017}} and consider a representation of the dynamics
(\ref{eq:eom-with-noise}) in discrete times $t_i=i\Delta t$,
with $i=0,1,2,\ldots$, and $\Delta t>0$ being a small time step. The
discrete-time dynamics is given by
\begin{equation}
\theta_i=\theta_{i-1}+\Delta t\left(\overline{F}(\theta_i)+\eta_i\right),
\label{eq:dynamics}
\end{equation}
where we have defined 
\begin{equation}
F(\theta_i)\equiv -V'(\theta_i),
\end{equation}
which for our model system~(\ref{eq:eom}) equals $F(\theta_i)=a \sin \theta_i-\sin 2\theta_i$ and
$\overline{F}(\theta_i)\equiv (F(\theta_{i-1})+F(\theta_i))/2$. In
writing Eq.~(\ref{eq:dynamics}), we have
used the Stratonovich rule \cite{Gardiner:2009} in discretizing the
dynamics (\ref{eq:eom}). The time-discretized Gaussian, white noise $\eta_i$
satisfies $\langle\eta_i\eta_j\rangle=\sigma^2\delta_{ij}$, where $\sigma^2$ is a positive constant with the dimension of $[{\rm
time-squared}]^{-1}$. In particular, the joint probability distribution
of occurrence of a given realization $\{\eta_i\}_{1\le i\le N}$ of the
noise, with $N$ being a positive integer, is given by 
\begin{equation}
P[\{\eta_i\}]=\left(\frac{1}{2\pi
\sigma^2}\right)^{N/2}\exp\left(-\frac{1}{2\sigma^2}\sum_{i=1}^{N}\eta_i^{2}\right).
\label{eq:joint-distribution}
\end{equation}

From the discrete-time dynamics (\ref{eq:dynamics}) and the joint distribution
(\ref{eq:joint-distribution}), the probability of
occurrence of a given phase trajectory $\{\theta_i\}_{0 \le i \le
N}\equiv\{\theta_0,\theta_1,\theta_2,\ldots,\theta_{N-1},\theta_N=\theta\}$
is obtained as
\begin{eqnarray}
&&P[\{\theta_i\}]={\rm det}({\cal J})\left(\frac{1}{2\pi
\sigma^2}\right)^{N/2}\prod_{i=1}^{N}\exp\left(-\frac{(\theta_i-\theta_{i-1}-\overline{F}(\theta_i)\Delta t)^2}{2\sigma^2(\Delta t)^2}\right).
\end{eqnarray}
Here, ${\cal J}$ is the Jacobian matrix for the transformation
$\{\eta_i\}\rightarrow\{\theta_i\}$, and is given by ${\cal J}_{1\leq i,j\leq
N}\equiv\left(\partial\eta_i/\partial \theta_j\right)$. For small
$\Delta t$, using ${\rm det}({\cal J})=(1/\Delta
t)^N\exp\left(-\sum_{i=1}^{N}(\Delta t/2)F'(\theta_i)\right)$, one gets by considering all possible
trajectories that the
probability density that the phase while starting at the value
$\theta_0$ at time $t=0$ evolves to the value $\theta$ at time
$t=N\Delta t$ is given by~\cite{Roldan:2017}
\begin{eqnarray}
&&P(\theta,t|\theta_0,0)=\left(\frac{1}{2\pi \sigma^2(\Delta
t)^2}\right)^{N/2}\prod_{i=1}^{N-1}\int_{-\pi}^\pi{\rm
d}\theta_i\nonumber \\
&&\times\exp\Big(-\Delta
t\sum_{i=1}^{N}\Big[\frac{[(\theta_i-\theta_{i-1}-\overline{F}(\theta_i)\Delta
t)/\Delta t]^{2}}{2\sigma^2 \Delta t}+\frac{F'(\theta_i)}{2}\Big]\Big).
\end{eqnarray}

In the limit of continuous time (i.e., $\Delta t \to 0$), using $D\equiv
\lim_{\sigma^2\to
\infty,\Delta t \to 0}(\sigma^2/2)\Delta t$, and defining
${\cal D}\theta(t)\equiv\lim_{N\to\infty}\Big(1/(4\pi D\Delta
t)\Big)^{N/2}\prod_{i=1}^{N-1}\int_{-\pi}^\pi{\rm
d}\theta_i$,  one gets an exact expression for the corresponding
probability density to be given by the following path integral~\cite{Roldan:2017}:
\begin{equation}
P(\theta,t|\theta_0,0)= \int_{\theta(0)=\theta_0}^{\theta(t)=\theta}{\cal
D}\theta(t)\exp\left(-S[\{\theta(t)\}]\right),
\label{eq:P-final}
\end{equation}
where we have introduced the action as
\begin{equation}
\hspace{-0.2cm}S[\{\theta(t)\}]=\int_0^{t}{\rm
d}t\left[\frac{[({\rm d}\theta/{\rm d}t)-F(\theta)]^{2}}{4D}+\frac{
F'(\theta)}{2}\right].
\end{equation}

We may now invoke the Feynman-Kac
formalism to identify the path integral on the
right hand side of Eq.~(\ref{eq:P-final}) with the propagator of a
quantum mechanical evolution in (negative) imaginary time due to a quantum
Hamiltonian $H_{\rm q}$. We then have
\begin{eqnarray}
P(\theta,t|\theta_0,0)&=&\exp\left(\frac{1}{2D} \int_{\theta_0}^\theta
F(\theta)~{\rm
d}\theta\right)G_{\rm q}(\theta,-it|\theta_0,0) \nonumber \\ 
&=&{\cal F}(\theta,\theta_0)G_{\rm
q}(\theta,-it|\theta_0,0),
\end{eqnarray}
with
\begin{eqnarray}
&&{\cal F}(\theta,\theta_0)\equiv \exp\left(\frac{1}{2D}
\left[a(\cos \theta_0-\cos \theta)+\frac{\cos 2\theta-\cos
2\theta_0}{2}\right]\right), \nonumber \\
\label{eq:qma} \\
&&G_{\rm q}(\theta,-it|\theta_0,0)\equiv\langle
\theta|\exp(-H_{\rm q}t)|\theta_0\rangle, \nonumber
\end{eqnarray}
where the quantum Hamiltonian is 
\begin{equation}
H_{\rm q}(\theta)\equiv-\frac{1}{2m_{\rm q}}\frac{\partial^{2}}{\partial
\theta^{2}}+V_{\rm q}(\theta),
\label{eq:quantum-hamiltonian}
\end{equation}
the mass in the equivalent quantum problem is 
\begin{equation}
m_{\rm q}\equiv\frac{1}{2D},
\end{equation}
and the quantum potential is given by
\begin{eqnarray}
&&V_{\rm q}(\theta)\equiv
\frac{(F(\theta))^2}{4D}+\frac{F'(\theta)}{2}=\frac{(a \sin \theta-\sin
2\theta)^2}{4D}+\frac{a\cos \theta-2\cos 2\theta}{2}.
\label{eq:quantum-potential}
\end{eqnarray}
Note that in the quantum propagator in Eq.~(\ref{eq:qma}), the Planck's
constant has been set to unity.

In terms
of the eigenvalues $E_n$ and the eigenfunctions $\Phi_n(\theta)$ of the
Hamiltonian $H_{\rm q}(\theta)$, we have
\begin{equation}
G_{\rm q}(\theta,-it|\theta_0,0)=\sum_n
\Phi_n(\theta)\Phi_n^\star(\theta_0)e^{-E_nt}.
\end{equation}
Hence, we have 
\begin{equation}
P(\theta,t|\theta_0,0)={\cal F}(\theta,\theta_0)\sum_n
\Phi_n(\theta)\Phi_n^\star(\theta_0)e^{-E_nt}.
\label{eq:general-1}
\end{equation}
In the limit $t \to \infty$, we may expect that only the eigenvalue equal to zero (provided it exists) will matter, so that we have
\begin{equation}
P(\theta,t \to
\infty|\theta_0,0)={\cal F}(\theta,\theta_0)\Phi_0(\theta)\Phi_0^\star(\theta_0).
\label{eq:larget-general}
\end{equation}
Equations~(\ref{eq:general-1}) and (\ref{eq:larget-general}) constitute our
exact expressions for the conditional probability. Obviously, the form
of the eigenvalues $E_n$ and eigenfunctions $\Phi_n$ depend on the form
of the potential $V(\theta)$.

\begin{figure}[!ht]
\centering
\includegraphics[width=9cm]{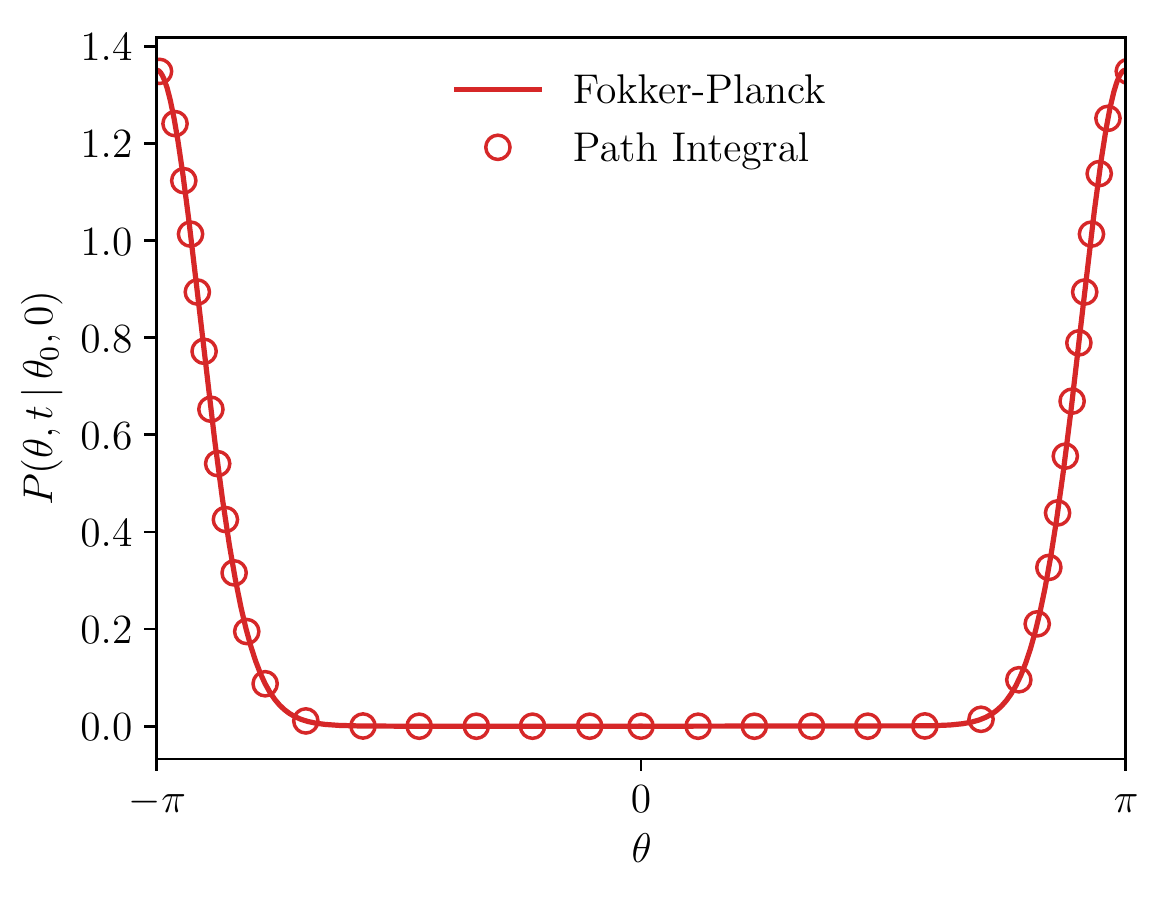}
        \caption{(Color online) Comparison of the conditional probability
        $P(\theta,t|\theta_0,0)$ obtained from the Fokker-Planck and
        the path integral approach. Here, we have chosen
        $a=4,D=0.5,t=5,\theta_0=0.75\pi$. Note that the probability is
        peaked at $\theta=\pm \pi$, the stable fixed point at this value
        of $a$, see Fig.~\ref{fig:bifurcation}.}
\label{fig:conditional-probability-comparison}
\end{figure}

In Fig.~\ref{fig:conditional-probability-comparison}, we show a
comparison of the conditional probability $P(\theta,t|\theta_0,0)$
obtained from the Fokker-Planck and the path-integral approach, for
representative values of $a,D,t$ and $\theta_0$. In the Fokker-Planck
approach, we have taken the truncation parameter to be $n_{\rm max}=60$,
making sure that higher values do not affect our results appreciably. In the path-integral
approach, we obtain the eigenvalues $E_n$ and the eigenfunctions
$\Phi_n$ of the
Hamiltonian~(\ref{eq:quantum-hamiltonian}) by discretizing $\theta$ over
$[-\pi,\pi]$, expressing the
Hamiltonian as a matrix and then solving numerically the corresponding eigenvalue
equation. Figure~\ref{fig:conditional-probability-comparison}
demonstrates an excellent agreement between the results obtained in the
two approaches.
From the figure, it
is evident that the probability is peaked at $\theta=\pm \pi$, the
stable fixed point at the considered value of $a$; our expectation is
that the density $P(\theta,t|\theta_0,0)$ gets more sharply
peaked as $D \to 0$, thus allowing to recover the stable fixed points of
the noiseless dynamics from the noisy one.

\begin{figure}[!ht]
\centering
\includegraphics[width=9cm]{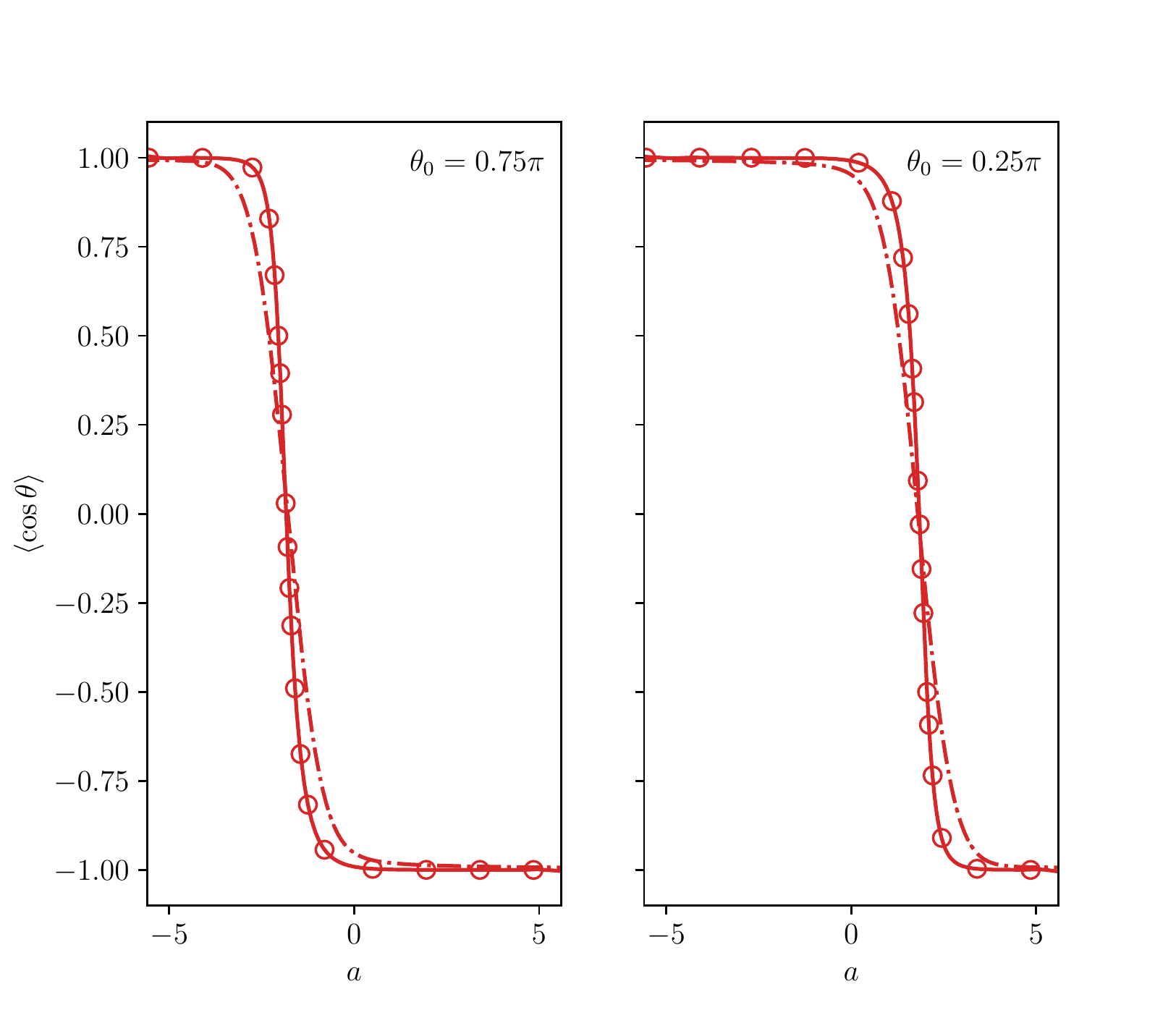}
\caption{(Color online) $\langle \cos \theta\rangle$ vs. $a$ at time
        $t=1$, and for
        $\theta_0=0.75\pi$ (left panel) and $\theta_0=0.25\pi$ (right panel). The lines
        are obtained by using
        Eq.~(\ref{eq:costheta-average-definition}): The continuous lines are for
        $D=10^{-5}$, while the dash-dotted lines are for $D=10^{-1}$. On the
        other hand, the circles are
        obtained by numerically integrating the noiseless
        dynamics~(\ref{eq:eom-D0}) with
        integration time step ${\rm d}t=10^{-3}$. The plots show that the
        curves corresponding to the noiseless dynamics coincide with the
        noisy ones in the limit $D \to 0$.}
\label{fig:vary-D-behavior}
\end{figure}

\section{Results and discussions}
\label{sec:results}

\begin{figure}[!ht]
\centering
\includegraphics[width=9cm]{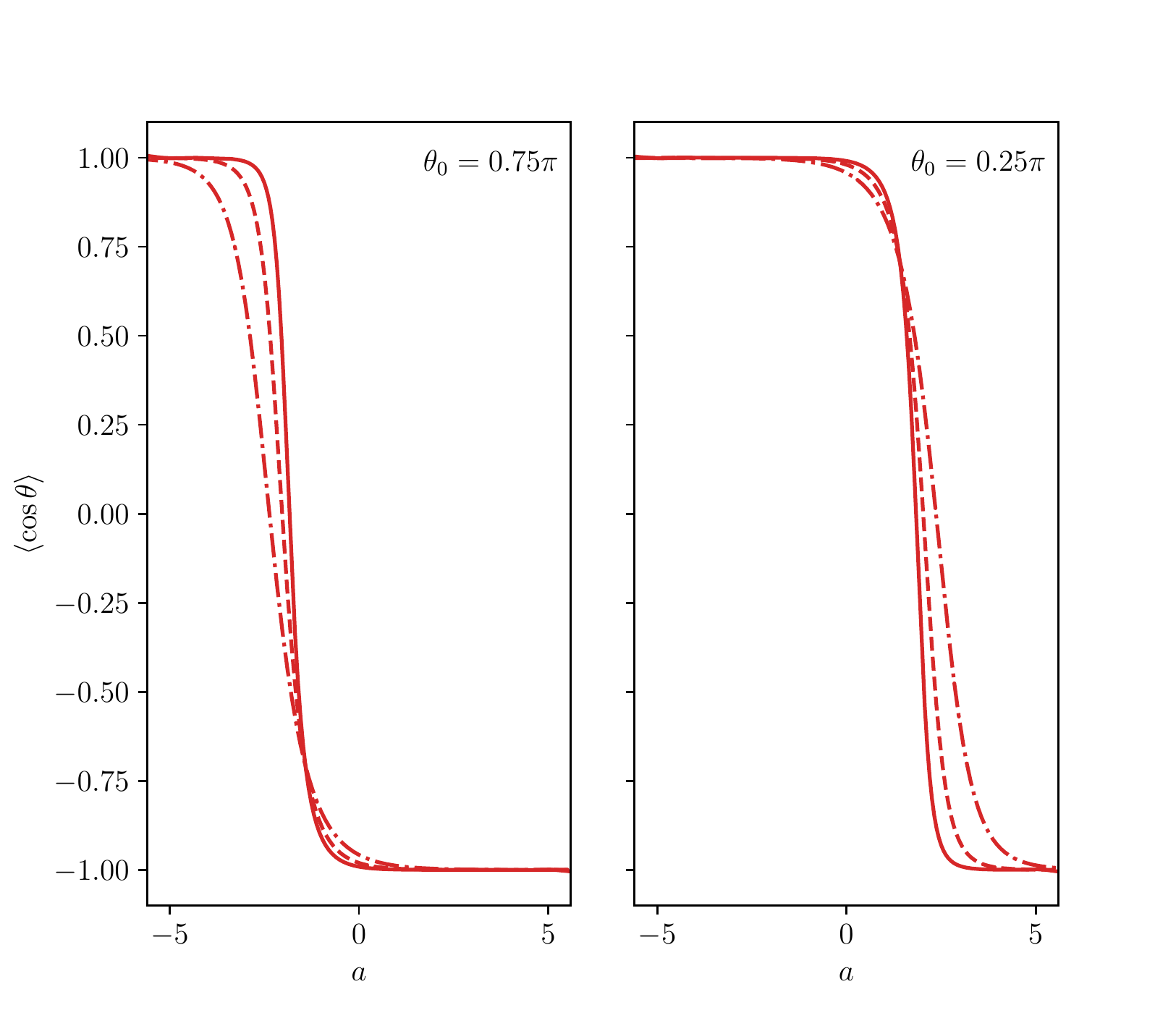}
\caption{(Color online) $\langle \cos \theta\rangle$ vs. $a$ for
        $\theta_0=0.75\pi$ (left panel) and $\theta_0=0.25\pi$ (right
        panel). The lines
        are obtained by using
        Eq.~(\ref{eq:costheta-average-definition}). Here, we have chosen
        $D=10^{-5}$. The dash-dotted, the dashed and the
        continuous line correspond respectively to times $t=0.6,0.8,1.0$. The plots show that the crossover between
        values $\langle \cos \theta\rangle=+1$ and  $\langle \cos
        \theta\rangle=-1$ with change of $a$ becomes steeper with the
        increase of $t$.}
\label{fig:small-D-behavior}
\end{figure}

We now discuss the results obtained from the analysis of the noisy
dynamics~(\ref{eq:eom}) discussed in the preceding section.
For a given initial value $\theta_0$ and a given noise strength $D$, we may calculate the average of $\cos \theta$ at a
given time $t$ and for different values of
$a$ by using either the Fokker-Planck or the path-integral result for
the conditional probability density $P(\theta,t|\theta_0,0)$, as
\begin{equation}
\langle \cos \theta \rangle \equiv \langle \cos \theta \rangle
(a,D,\theta_0,t)=\int_{-\pi}^\pi {\rm d}\theta~\cos \theta
P(\theta,t|\theta_0,0).
\label{eq:costheta-average-definition}
\end{equation}
Figure~\ref{fig:vary-D-behavior} shows that the results obtained in the limit $D \to 0$ (specifically, for
$D=10^{-5}$) are in excellent agreement with the values of $\cos \theta$
estimated from numerical integration of the noiseless
dynamics~(\ref{eq:eom-D0}). This is consistent with Fig.~\ref{fig:space-time-trajectory} showing that
the noisy trajectories in the limit $D \to 0$ represent small
fluctuations about the trajectories obtained in the noiseless dynamics. 

\begin{figure}[!ht]
\centering
\includegraphics[width=9cm]{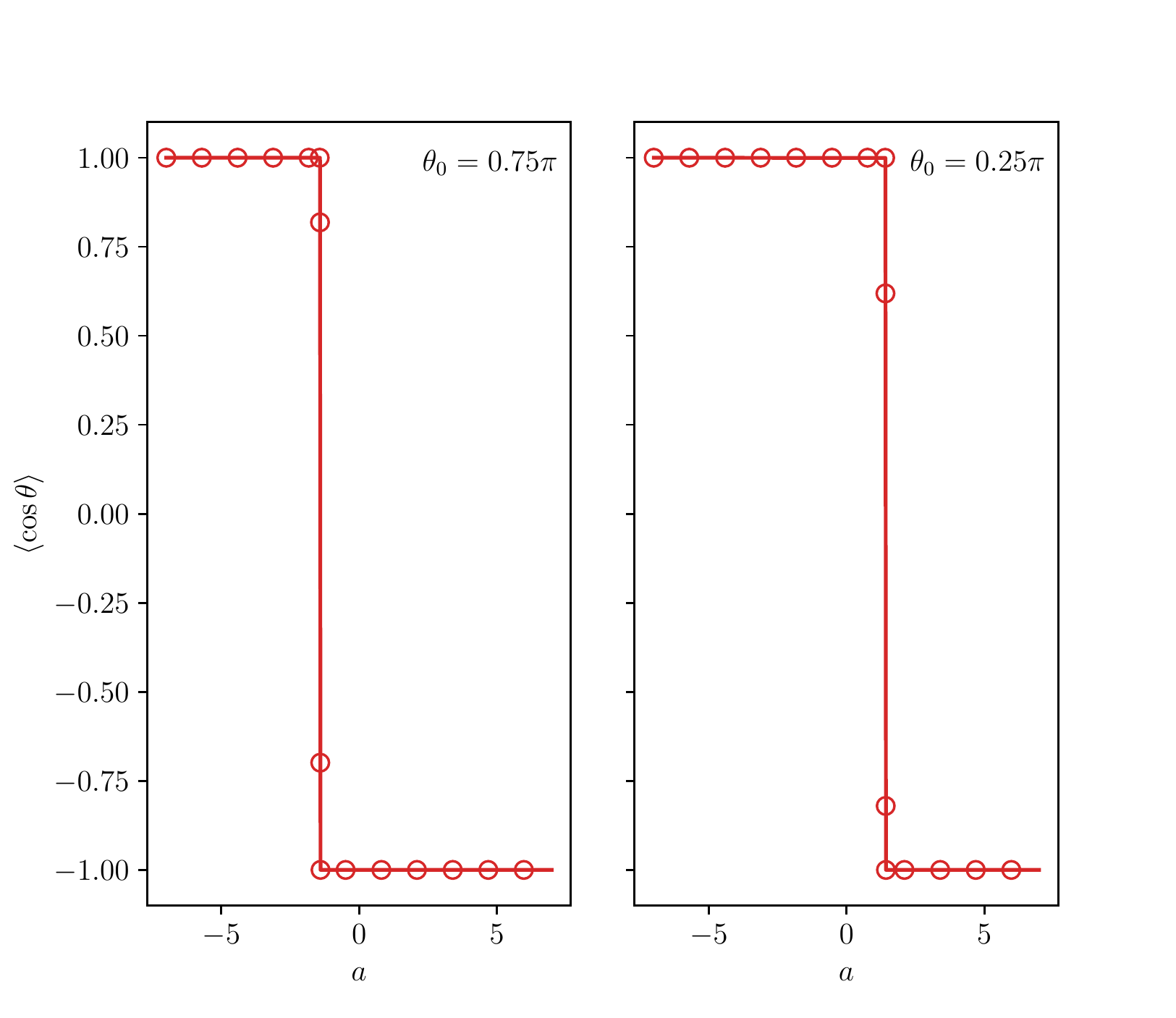}
\caption{(Color online) $\langle \cos \theta\rangle$ vs. $a$ obtained in
the stationary state (time $t=10$). The left panel
        (respectively, the right panel) corresponds to the initial condition $\theta_0=0.75\pi$
        (respectively, $\theta_0=0.25\pi$). Here, we have $D=10^{-5}$. While the red circles
        involve using Eq.~(\ref{eq:costheta-average-definition}), the lines are obtained from
        numerical integration of the noiseless
        dynamics~(\ref{eq:eom-D0}) using integration time step ${\rm d}t=10^{-3}$.}
\label{fig:small-D-steady-state-behavior}
\end{figure}

In the next step towards obtaining the bifurcation behavior of the
noiseless dynamics from the noisy one, we take $D=10^{-5}$, and
obtain for a given $\theta_0$ and a given time $t$ the behavior of
$\langle \cos \theta \rangle$ versus $a$ by using
Eq.~(\ref{eq:costheta-average-definition}). It is evident from the results shown in
Fig.~\ref{fig:small-D-behavior} that the
cross-over between the two limiting values of $\langle \cos \theta\rangle$,
namely, $\langle \cos \theta \rangle=+1$ and $\langle \cos \theta \rangle=-1$, becomes
steeper with the increase of $t$. Indeed, for larger $t$, one has a
sharp jump, as shown in Fig.~\ref{fig:small-D-steady-state-behavior}.
The same results are obtained as $t$ is increased further, so
Fig.~\ref{fig:small-D-steady-state-behavior} characterizes stationary behavior.  In
this case, we further show that the long-time values of $\cos \theta$
obtained in the noiseless dynamics lie on the curve for the noisy
dynamics. 

Referring to Fig.~\ref{fig:bifurcation}, for a
given $\theta_0$, consider increasing $a$ from low to high values, that
is, moving along a straight line parallel to the $x$-axis and at a
distance $\theta_0$ from it. Then, with change of $a$, the
long-time value of $\cos \theta$ in the noiseless dynamics will be $+1$ so long as the straight
line does not intersect the dashed curve in red lying in the region
$-2 < a <2$. Beyond the point of intersection, the long-time value of
$\cos \theta$ will be $-1$. The point of intersection, obtained
by solving $\cos \theta_0=a/2$, will thus be a crossover point such that
for smaller (respectively, larger) $a$, the stable value of $\cos \theta$ will
be $+1$ (respectively, $-1$). In view of Fig.~\ref{fig:vary-D-behavior} showing match
between the noiseless and the noisy dynamics in the limit of weak noise, such a behavior would be
expected of $\langle \cos \theta \rangle$ versus $a$ at long times and is indeed borne
out by our exact results shown in
Fig.~\ref{fig:small-D-steady-state-behavior}. It may be checked from the
figure that the crossover point is obtained at the value of $a$ given by
$a=2\cos \theta_0=\pm \sqrt{2}$ for the left and the right panel,
respectively.  Obtaining the crossover
point by repeating plots as in
Fig.~\ref{fig:small-D-steady-state-behavior} for different values of
$\theta_0$ allows to obtain the
line of unstable fixed points in the range $-2 < a < 2$. In
Fig.~\ref{fig:bifurcation}, we show that as expected, the crossover
points so obtained lie exactly on the unstable branch in the range $-2 <
a < 2$. Repeating plots as in
Fig.~\ref{fig:small-D-steady-state-behavior} for $\theta_0=0$ and
$\theta_0=\pi$ allows to obtain the crossover points $a=2$ and $a=-2$,
respectively. These points coincide with the bifurcation points in
Fig.~\ref{fig:bifurcation}, thereby explaining the associated stability.

\begin{figure}[!ht]
\centering
\includegraphics[width=9cm]{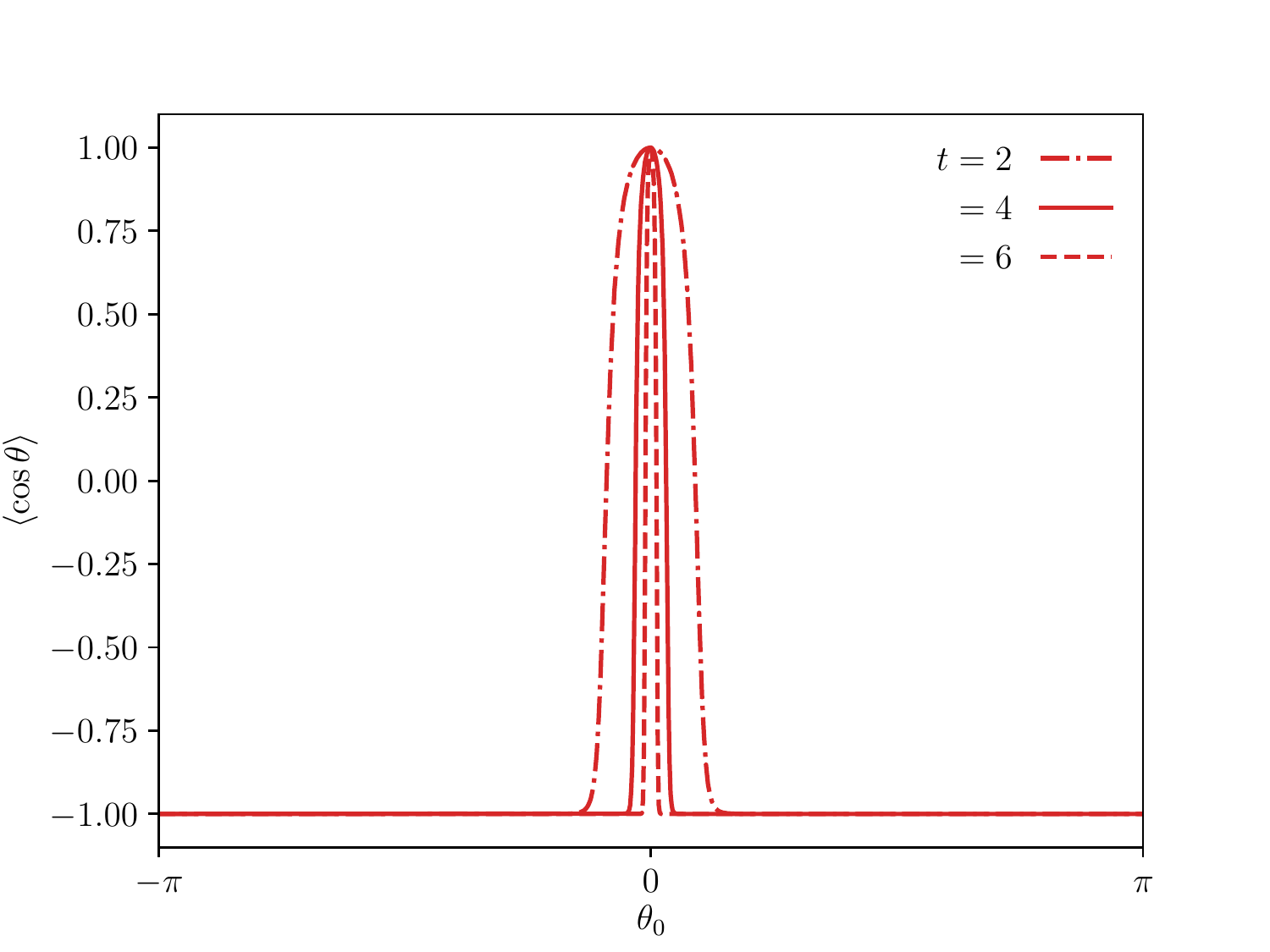}
\caption{(Color online) $\langle \cos \theta\rangle$ vs. $\theta_0$
obtained using Eq.~(\ref{eq:costheta-average-definition}) and for
$a=2.5$. Here, we have $D=10^{-5}$, and the different curves correspond to
different times $t$.}
\label{fig:costheta-a2-5}
\end{figure}

The unstable fixed points of the noiseless dynamics may also be obtained
from the noisy dynamics. For example, in order to arrive at the fact that $\theta=0$
is unstable at $a=2.5$ (see Fig.~\ref{fig:bifurcation}), one may plot $\langle \cos \theta \rangle$,
obtained using Eq.~(\ref{eq:costheta-average-definition}), as a function
of $\theta_0$ and for different times.
From Fig.~\ref{fig:costheta-a2-5}, one may observe that as time
increases, $\langle \cos \theta \rangle$ for increasing number of values
of $\theta_0$ different from the specific value $\theta_0=0$ attains the
value of $-1$. From the curves for different times, it is evident that in
the limit of long times, only when $\theta_0=0$ does $\langle \cos
\theta \rangle$ have the value of unity, which attains for all other values of
$\theta_0$ the value of $-1$. This is fully consistent with the
bifurcation diagram~\ref{fig:bifurcation}, and has been indicated in the
figure by the red circle at $\theta=0,a=2.5$.

\section{Conclusions}
\label{sec:conclusions}
In this work, we addressed the issue of how one may obtain the
bifurcation behavior of a non-linear dynamical system by introducing
noise into the dynamics and then studying the resulting Langevin
dynamics in the weak-noise limit.
Within the ambit of a model system, we showed that a suitable quantity to
capture the bifurcation behavior in the noisy dynamics is to define a conditional probability to observe microscopic configurations
at a given time while conditioned on observation of a given
configuration at an earlier time. The time evolution of the conditional
probability may be studied by using two complementary approaches,
namely, the Fokker-Planck and the path-integral approach, with the
latter yielding exact closed-form expressions for the conditional
probability. 

The analysis presented in
Section~\ref{sec:analysis-noisy-dynamics} applies to any potential
$V(\theta)$, and therefore, the whole program of obtaining the
bifurcation diagram of a given noiseless dynamics corresponding to a given form of
$V(\theta)$ by addition of Gaussian, white noise to the dynamics can be rather straightforwardly
carried through. A remarkable feature of the latter approach is that the
probability distribution $P(\theta,t|\theta_0,0)$ obtained from either
the Fokker-Planck or the path-integral approach is for weak-enough noise
and at long times 
naturally peaked around the stable fixed points of the noiseless
dynamics. In this way, once for a given noiseless
dynamics one obtains its fixed points, one may bypass the need to
perform a stability analysis of the fixed points in order to locate the
stable ones, by studying the
corresponding noisy dynamics in the limit of weak noise. 

A reason why we could recover the bifurcation behavior
in the noisy dynamics is the choice of Gaussian noise for Langevin
evolution, which ensures that typical trajectories for the noisy
dynamics represent fluctuations of a given size (set by the variance $D$ of
the Gaussian distribution for the noise) around the noiseless ones, and hence
coincide with the latter in the limit $D \to 0$. A question that
naturally arises in this regard is: numerically how small should $D$ be?
The answer depends on whether
one is studying the dynamics around a stable or an unstable fixed point.
(a) For a stable fixed point, which corresponds to a local minimum of the
potential $V(\theta)$, any reasonably small value of $D$ (smaller than a
critical value $D=D_c^{(s)}$) would ensure
that almost all trajectories of the noisy dynamics are pushed towards
the fixed point by virtue of the potential having a minimum at the
stable point, thereby settling into the fixed point at long times. (b) For an unstable fixed point, the issue is a bit tricky.
Since such a fixed point corresponds to a local maximum of $V(\theta)$,
values of $D$ smaller that $D_c^{(s)}$ that ensured convergence of the
noisy to the noiseless dynamics in (a) may prove to be ``strong" enough
that trajectories starting at the unstable fixed point are pushed away
from it to settle into stable fixed points at long times. This would at
once invalidate our procedure of obtaining the unstable fixed points of
the noiseless dynamics from an analysis of the noisy one. In order that the program is successful, one would be
required to reduce further the noise strength, i.e., having a critical
value $D_c^{(u)} < D_c^{(s)}$, and considering for convergence values of
$D < D_c^{(u)}$. 

\begin{figure}[!ht]
\centering
\includegraphics[width=9cm]{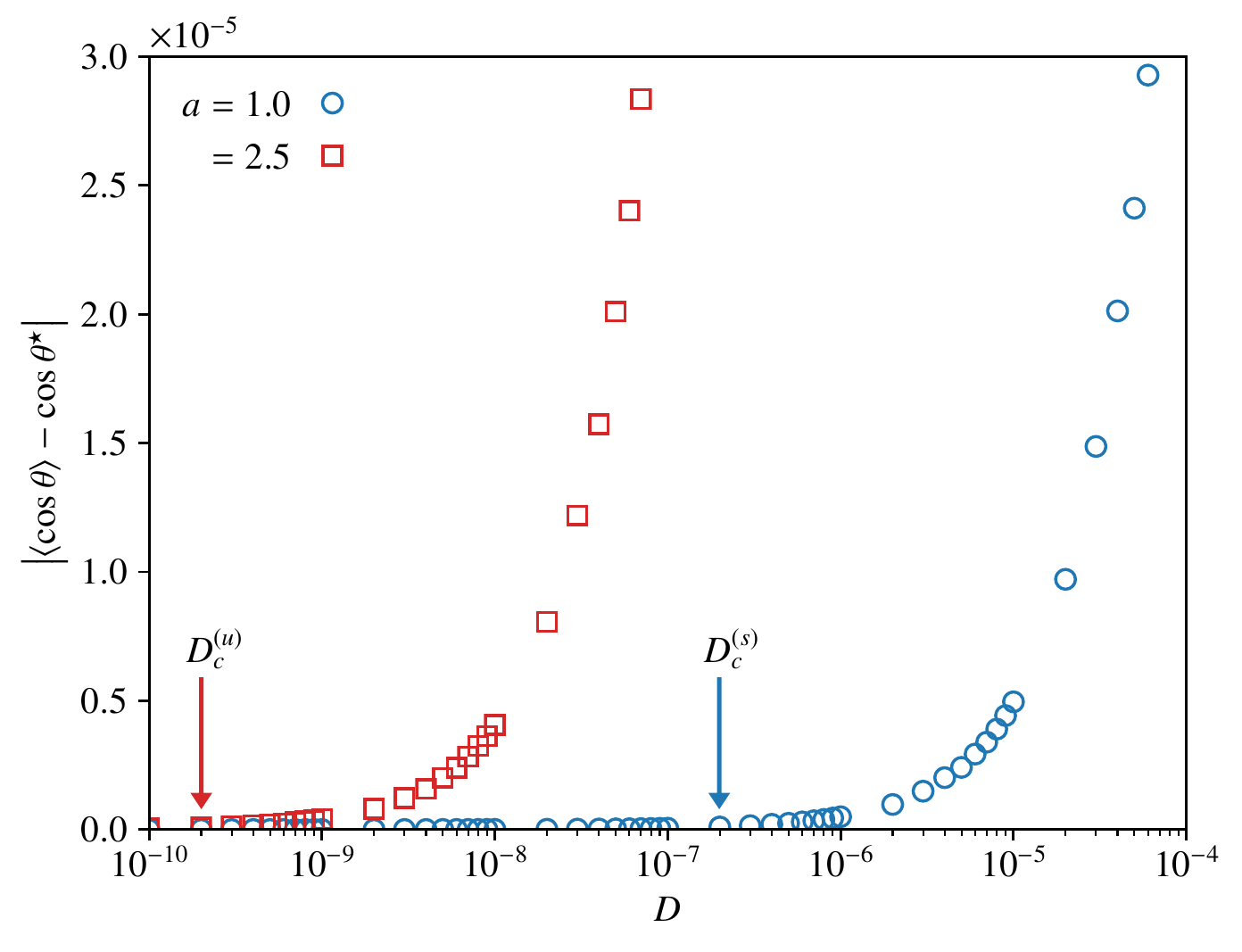}
\caption{(Color online) $|\langle \cos \theta\rangle-\cos \theta^\star|$
vs. $D$ for values of $a$ at which one has a fixed point at
$\theta=\theta^\star$. For $a=1.0$ (respectively, $a=2.5$), one has a
stable (respectively, unstable) fixed point, with $\theta^\star=0$ for
both. Here, one obtains $\langle \cos \theta\rangle$ by using
Eq.~(\ref{eq:costheta-average-definition}) and by choosing
$\theta_0=\theta^\star$ and $t=6$. In the figure, we have indicated the
approximate $D_c^{(s)}$ and $D_c^{(u)}$.}
\label{fig:D-estimate}
\end{figure}

To
illustrate that the aforementioned conclusion is indeed borne out by our
results, we show in Fig.~\ref{fig:D-estimate} the outcome of the following numerical experiment. Referring to
Fig.~\ref{fig:bifurcation}, we choose a value of $a$ at
which one has a stable fixed point $\theta^\star$, and another at which one has an
unstable fixed point $\theta^\star$, and study the noisy dynamics at these values
of $a$ and with the initial value $\theta_0=\theta^\star$. For both, we expect that for $D$ small enough
(i.e,, for $D< D_c^{(s)}$ (respectively, $D < D_c^{(u)}$) for the stable
(respectively, the unstable) fixed point), the long-time value of $\langle \cos \theta
\rangle$ should coincide with the value of $\cos \theta^\star$. From the
figure, we see that indeed we have $D_c^{(u)} < D_c^{(s)}$. Of course, the precision of match between
the values of $\langle \cos \theta \rangle$ and $\cos \theta^\star$ depends on
the precision employed in numerical evaluation of the quantities involved, and higher precision implies lower values of $D_c^{(s)}$ and
$D_c^{(u)}$, a fact
we have checked in our numerics.

We wrap off by mentioning an utility of studying the noisy dynamics. Any
modeling of experimental data on the behavior of a real system by a dynamics should
include effects of noise to account for measurement errors. The actual
underlying dynamics of course does not have this source of noise, and one is typically interested in inferring how
qualitatively different behavior observed in the data obtained with varying
experimental parameters emerges from a bifurcation in the actual
dynamics. A long-time analysis of the noisy dynamics automatically
picks up the stable points of the actual dynamics (experimental data
typically contain signatures of only stable points, with unstable points
contributing mainly to short-time transients), thus allowing to infer directly the
bifurcation behavior of the actual noiseless system from a study of the noisy
dynamics.

As a concrete application of our method, we may
mention the following scenario: Systems of neurons exhibit diverse dynamical
behaviors depending on values of biophysical parameters, such as
quiescence, spiking, bursting, and many others. A phenomenological
neuron model proposed by Hindmarsh and Rose (the HR
model~\cite{HR:1984}) is known to numerically
exhibit all of the above behaviors. Bifurcations in the model as one
tunes the various dynamical parameters have been studied mostly
numerically or under suitable approximations in specific parameter
regimes, owing to challenges
involved in pursing a complete analytical study of the model, see
Ref.~\cite{Storace:2008} for a recent study. It will be
interesting to see whether adding noise to the model and studying the
resulting noisy dynamics in the weak-noise limit allows to obtain the
complete bifurcation diagram of the HR model. It is
also left for future work as to how one may extract other features of dynamical
systems by studying the corresponding noisy dynamics that allows to use
standard tools of statistical physics, e.g., the Fokker-Planck and the
path-integral approach.

\section{Acknowledgements}
The work of Debraj Das is supported by UGC-NET Research Fellowship Sr.
No. 2121450744, dated 29-05-2015, Ref. No. 21/12/2014(ii) EU-V.
Sayan Roy acknowledges DST-INSPIRE, Government of India for providing
him with a scholarship to do a summer project at the Ramakrishna Mission
Vivekananda University during May -- July, 2018. 




\begin{thebibliography}{99}
\bibitem{note-euler}Unless stated otherwise, in this work we use the Euler scheme to perform numerical integration of
equations of motion.
\end{thebibliography}
\end{document}